\documentclass[letter,10pt,5p,twocolumn]{elsarticle}
\pdfoutput=1

\usepackage{lineno}
\usepackage{subcaption}
\usepackage{graphicx}
\journal{Physics Letter B}

\begin{document}

\begin{frontmatter}

\title{Acceleration by Strong Interactions}

\author{M.~Erdmann}
\ead{erdmann@physik.rwth-aachen.de}
\author{C.~Glaser}
\author{T.~Quast}
\address{Physics Institute 3A, RWTH Aachen University, D-52056 Aachen, Germany}
\begin{abstract}
Beyond the attractive strong potential needed for hadronic bound states,
strong interactions are predicted to provide repulsive forces
depending on the color charges involved.
The repulsive interactions could in principle serve for particle acceleration 
with highest gradients in the order of GeV/fm.
Indirect evidence for repulsive interactions have been reported in the context
of heavy meson production at colliders.
In this contribution, we sketch a thought experiment to directly investigate 
repulsive strong interactions.
For this we prepare two quarks using two simultaneous deep inelastic 
scattering processes off an iron target.
We discuss the principle setup of the experiment and estimate the 
number of electrons on target required to observe a repulsive effect between the quarks.
\end{abstract}


\end{frontmatter}

\section{Motivation}

In order to access smallest structures, the quest for high energies is mandatory. 
It is commonly believed that corresponding accelerator sites need to be large, 
and that even cosmic rays with $10^{20}$~eV are accelerated by electromagnetic 
interactions in giant cosmic structures. 
If, instead, strong interactions with gradients in the order of GeV/fm could be exploited, acceleration 
sites could in principle fit the size of a laboratory experiment. 

\begin{figure}[htb]
\centering
 \includegraphics[width=0.5\textwidth]{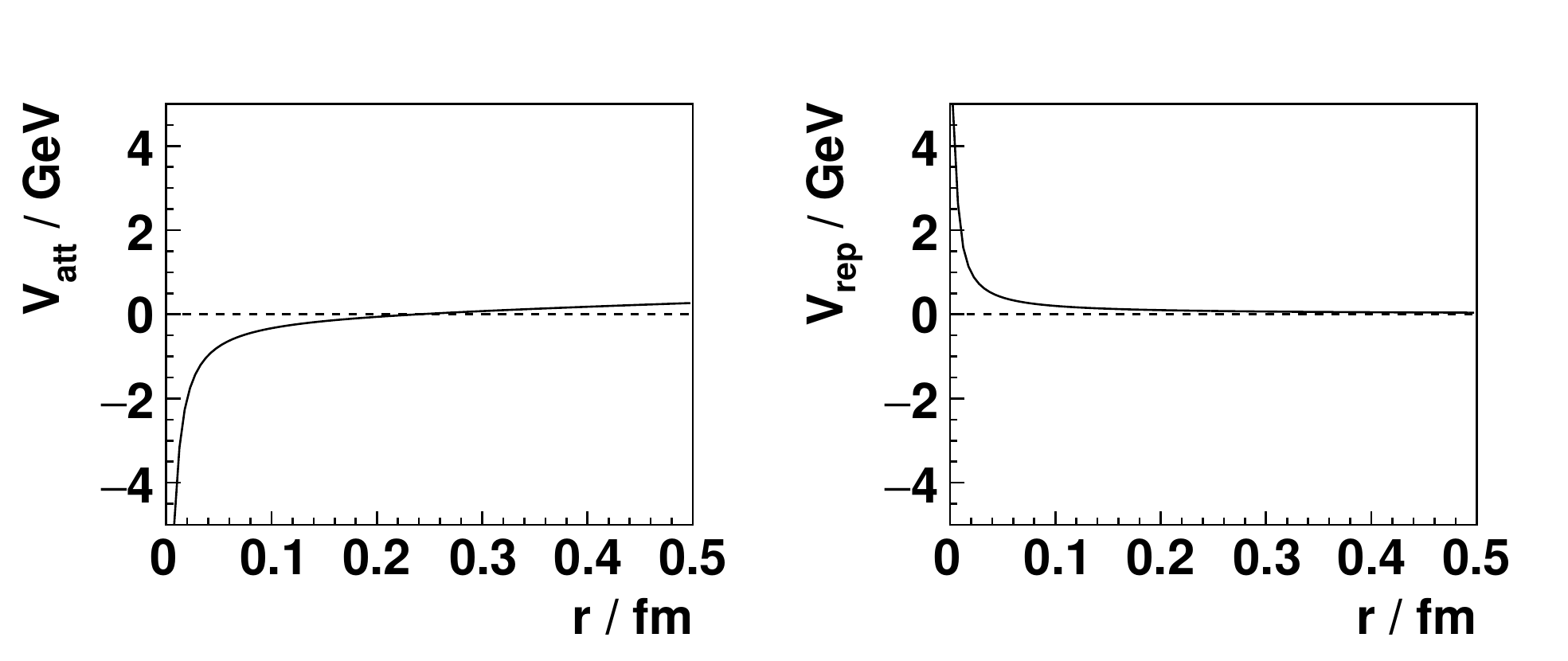}
\begin{minipage}[]{.23\textwidth} \centering{(a)} \end{minipage}
\begin{minipage}[]{.23\textwidth} \centering{(b)} \end{minipage}
 \caption{Strong interaction potential, a) attractive, b) repulsive.}
\label{fig:potential}
\end{figure}

Different aspects related to acceleration by strong interactions have been investigated before.
For example, in fixed-target lepton-proton experiments, target protons were found to be accelerated 
up to GeV energies \citep{arneodo}.
The distance dependence of the attractive strong potential $V$ for quark-antiquark states 
(Fig.~\ref{fig:potential}a) has been studied in onium experiments 
\citep{eichten_theory, eichten_experiment, quigg, bali, hagler, ayala}, and has been investigated 
in deceleration processes of quarks and antiquarks in $Z$ boson decays \citep{erdmann1}. 

For combinations of quark-antiquark pairs in color singlet states or two quarks 
in color triplet states, an attractive force is expected from color factors as 
predicted by Quantum Chromodynamics (QCD) \citep{QCD} which is supplemented by
a long-range term representing quark confinement.
The attractive potential for color singlets reads
\begin{equation}
V_{att} = - \frac{4}{3} \frac{\alpha_s\;\hbar c}{r} + \kappa \; r \; .
\label{eq:attractive}
\end{equation}
Here, $r$ is the distance between the quarks, $\alpha_s$ is the strong coupling constant,
and $\kappa$ represents the strength of quark confinement.
$\hbar$ and $c$ refer to the Planck constant and the vacuum velocity of light, respectively.

Instead, for combining two quarks in a color sextet state or quark-antiquark pairs in color
octet states a repulsive force (Fig.~\ref{fig:potential}b) is expected from the corresponding
color factors of QCD \citep{QCD}. 
The repulsive potential for a color sextet state is denoted by
\begin{equation}
V_{rep} = \frac{1}{3} \frac{\alpha_s\;\hbar c}{r} \; .
\label{eq:repulsive}
\end{equation}

In principle, these repulsive forces could serve for acceleration purposes. 
Indirect evidence for repulsive color octet contributions has been reported in heavy meson 
physics at colliders before \citep{sansoni, braaten, bodwin}. 
A repulsive strong force, however, has not yet been measured directly. 

Building an accelerator site based on strong interactions is obviously a major, long-term project. 
The challenges call for a decade of research, as preparation of quarks, gluons, or generally 
colored objects is required. 
In addition, ways to control the strong potential and to concatenate 
acceleration phases need to be found. 

As a first step towards exploiting particle acceleration by strong interactions, 
we study a thought experiment for direct observation of the repulsive force between two quarks. 
The pathbreaking challenge to be solved is direct control of the two quarks.

\begin{figure}[ttt]
\centering
 \includegraphics[width=0.45\textwidth]{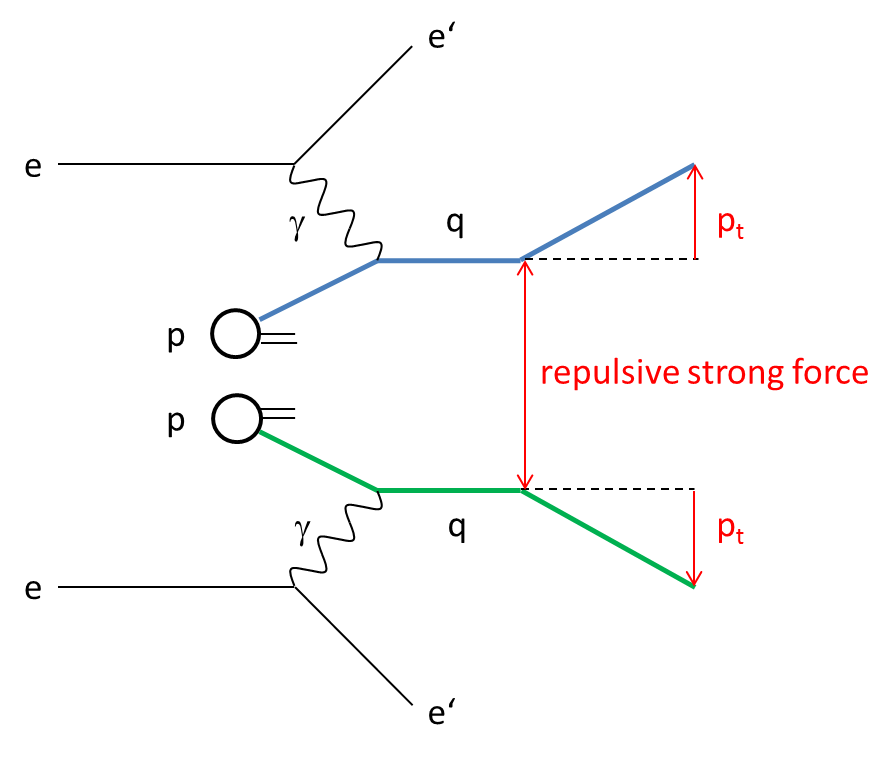}
 \caption{ Two deep inelastic lepton-nucleon scattering processes where the two scattered quarks 
repel each other owing to strong interactions.
}
 \label{fig:sketch}
\end{figure}

\begin{figure}[ttt]
\centering
 \includegraphics[width=0.45\textwidth]{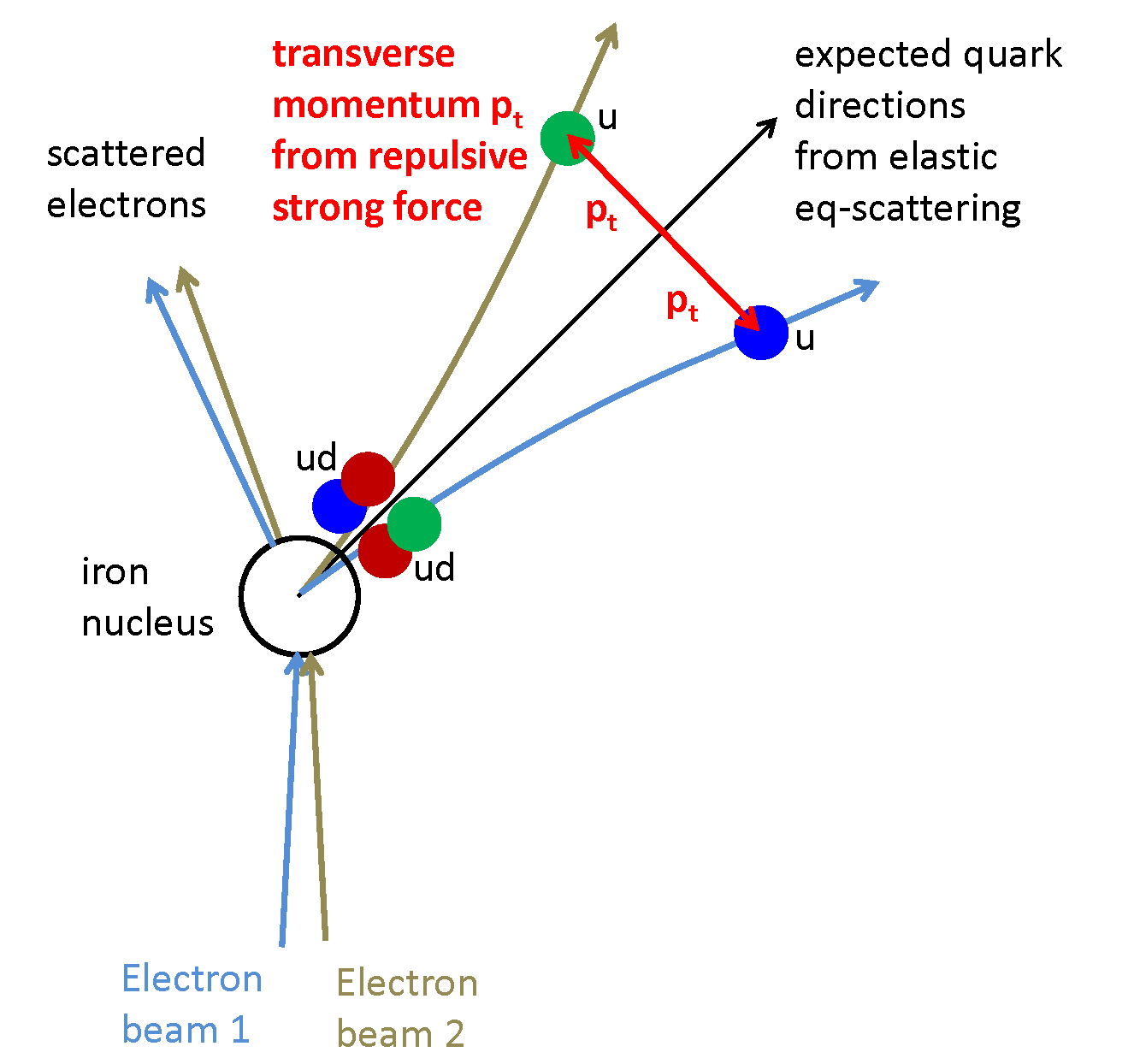}
 \caption{
Principle of the experimental setup. 
Two electron beams are directed to the fixed target area.
Candidate events have two scattered electrons such that the 
expected scattered quark directions coincide.
Repulsive strong interactions are indicated by hadronic 
final states exhibiting transverse momenta $p_t$ with respect to the
expected quark directions.
}
 \label{fig:setup}
\end{figure}

Preparation of two quarks can be achieved by two deep inelastic scattering processes off a nucleus
(Fig.~\ref{fig:sketch}). 
Two electron beams are directed, for example, at a low-degree angle onto an iron target (Fig.~\ref{fig:setup}).
The angles and energies of the scattered electrons determine the energies and flight directions 
of the two quarks.
In this way, events with the two quarks having a similar direction can be selected 
such that the quark-quark center-of-mass energy $\sqrt{s_{qq}}$ is small. 
A low $\sqrt{s_{qq}}$ is needed in order to avoid generating transverse momentum
from standard QCD quark-quark scattering processes.


If a repulsive force between the two quarks exists as predicted by QCD, the two quarks will 
acquire transverse momentum relative to their original flight direction.
The quark transverse momenta will be measured as hadron transverse momenta in an appropriate 
detector system (Fig.~\ref{fig:setup}).

In the following, we will first estimate the simultaneous cross-section for two deep inelastic 
scattering processes off a single iron nucleus.
We will then estimate the transverse momenta resulting from the repulsive interactions.
Finally, we will estimate the number of events obtained with an assumed experimental setup.

\section{Probability of two electrons scattering off an iron nucleus}

To estimate the probability of two electrons scattering off a single iron nucleus,
our strategy is to fold the well-known deep inelastic lepton-nucleon cross-section 
with a second lepton interaction.
To keep the quark-quark center-of-mass energy low, the second interaction will be 
required to have a quark scattering direction close to the direction of the other 
scattered quark.

\subsection{Cross-section for a single electron - single iron nucleus scattering}

\begin{figure}[htb]
\centering
 \includegraphics[width=0.45\textwidth]{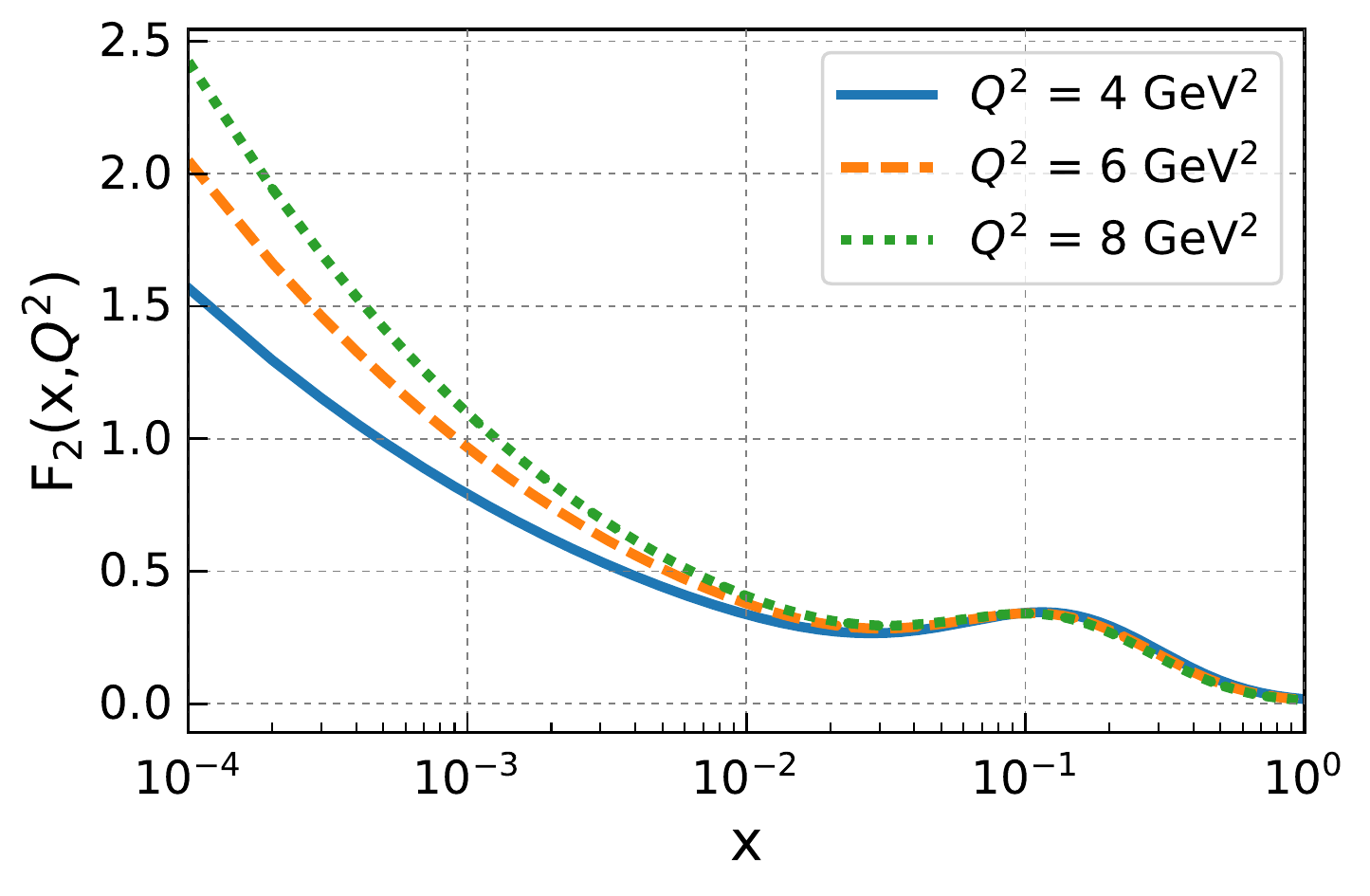}
 \caption{Structure function of the proton used in the calculation of the 
deep inelastic electron-iron scattering.}
 \label{fig:F2}
\end{figure}

The double differential cross-section for deep inelastic electron-iron scattering reads \citep{pdg}:
\begin{eqnarray}
  \frac{d^2\sigma}{dx\, dQ^2} &=& A^{2/3}\;
\frac{\alpha^2 ( 1+ (1-Q^2/s)^2 )}{8 \pi\,x\,Q^4} \;    F_2(x,Q^2) 
\label{eq:cross section}
\end{eqnarray}
Here, $Q^2$ is the squared four-momentum transfer by the exchanged photon, and 
$x$ denotes the parton fractional momentum, often referred to as Bjorken-$x$.
The term $A^{2/3}$ is the enhancement factor for the iron nucleus with $A=56$, resulting in
$A^{2/3} \approx 15$ where we neglect the EMC effect \citep{emc}.
As $Q^{2}$ is small compared to $s$ for most of the events, we approximate the term in 
brackets resulting from spin $1/2$ scattering by 
$(1+ \left(1-Q^2/s\right)^2) \approx 2$.
As the structure function, we use a parameterized version following the phenomenological ansatz
$F_2(x,Q^2)=a(x)\left[\ln{(Q^2/\Lambda^2)}\right]^{\kappa(x)}$ \citep{erdmann2}
with the result shown in Fig.~\ref{fig:F2}.

To enable similar directions of the two scattered quarks, we first replace the 
fractional momentum $x$ by the scattering angle $\theta_q$ of the quark.
The pseudorapidity $\eta_q=-\ln\tan(\theta_q/2)$ of the scattered quark in the electron-proton 
center-of-mass frame can be expressed by the squared four-momentum transfer $Q^2$ and the 
quark fractional momentum $x$.
At electron - proton colliders the angle is \citep{aid}:
\begin{equation}
 \eta_q(x,Q^2) =\frac{1}{2}
                \ln\left(\frac{x\, E_p}{E_e} \left(\frac{x\,s}{Q^2} -1\right)\right) \, .
\label{eq:quark_rapidity}
\end{equation}
Here, $E_e$ is the electron beam energy, and $E_p$ is the proton energy
for which we use the proton rest mass.

\begin{figure}[htb]
\hspace*{1cm} a) \hspace*{1cm} 
 \includegraphics[width=0.17\textwidth]{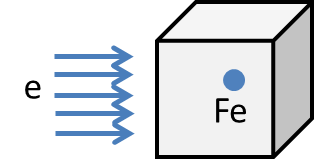}\\
\vspace*{0.5cm} \\
\hspace*{1cm} b) \hspace*{1cm} 
 \includegraphics[width=0.17\textwidth]{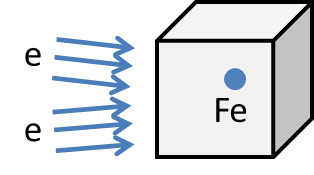}
\vspace*{0.5cm} \\
\hspace*{1cm} c) \hspace*{1cm} 
 \includegraphics[width=0.17\textwidth]{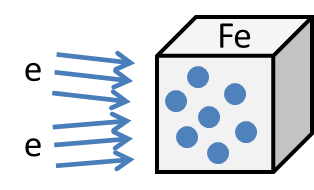}
 \caption{a) Electron beam on a single iron nucleus.
b) Two electron beams on a single iron nucleus.
c) Two electron beams on an iron target.}
 \label{fig:cube}
\end{figure}

Using analytical tools, eq.~(\ref{eq:quark_rapidity}) is converted to calculate the
quark fractional momentum $x(\theta_q, Q^2)$ as a function of the quark scattering
angle $\theta_q$ and $Q^2$.
In the lepton-iron cross-section (\ref{eq:cross section}) we replace the quark fractional 
momentum $x$ by $x(\theta_q,Q^2)$:
\begin{eqnarray}
 \frac{d^2\sigma}{d\theta_q dQ^2} &=& A^{2/3}\;\frac{\alpha^2}{4 \pi}\, \frac{1}{Q^4}\, 
\frac{1}{x(\theta_q,Q^2)}\, \;\;\;\;\;\;  \nonumber \\
&& F_2(x(\theta_q, Q^2),Q^2)\, 
  \left\vert\frac{dx}{d\eta_q}\right\vert \left\vert\frac{d\eta_q}{d\theta_q}\right\vert 
\label{eq:xsec_rapidity}
\end{eqnarray}
Integrating this differential cross-section above, $Q^2=Q^2_\circ$
gives an estimate of the total cross-section for deep inelastic electron scattering 
off a single iron nucleus as visualized in Fig.~\ref{fig:cube}a.
To ensure deep inelastic scattering processes we chose $Q^2_\circ=3.5$~GeV$^2/c^2$.

\subsection{Cross-section for two electrons - single iron nucleus scattering}

Here, we want to obtain the total cross-section of two electrons interacting with the same iron nucleus 
(Fig.~\ref{fig:cube}b) which then produce two quarks flying into approximately the same direction.
For this we calculate the cross-section of the first electron producing a 
quark flying in a specific direction and multiply this with the cross-section of the second 
electron producing a quark flying in the same direction. 

To obtain this product of cross-sections, we simultaneously integrate differential cross-sections 
for two electrons, each from one of the beams, where we demand 
the quark scattering angles to be within $\delta = \pm 10$~deg.
This is to ensure that the two scattered quarks move in the same direction and have a low
quark-quark center-of-mass energy.
For an electron beam energy of $E_e = 60$~GeV and a typical quark fractional momentum $x\sim 0.1$,
the squared quark-quark center-of-mass energy with $\delta=10$~deg amounts to:
\begin{equation}
s_{qq} = (p_{e1}+p_{e2})^2 \approx x_1 x_2 E_e^2 (1-\cos{\delta}) \approx (0.1~\rm{GeV})^2
\label{eq:sqrtsqq}
\end{equation}
Therefore, $\sqrt{s_{qq}}$ is low and does not lead to significant transverse momentum from
standard QCD quark-quark scattering processes.

The cross-section is then integrated above $Q^2\ge Q^2_\circ=3.5~\rm{GeV}^2/c^2$, and with the constraint
on the scattering angle of the second quark.
For the cross section of the second electron we omit the factor $A^{2/3}$ in eq. (\ref{eq:xsec_rapidity})
to have the two interactions close to each other.
\begin{eqnarray}
\tilde{\sigma}^2=2\cdot \int_{\theta_{q1}=0^\circ}^{\theta_{q1}=180^\circ} \int_{\theta_{q2}=\theta_{q1}-10^\circ}^{\theta_{q2}=\theta_{q1}+10^\circ} \int_{Q^2_{1}=Q^2_\circ}^{Q^2_{1}=E_e^2} \int_{Q^2_{2}=Q^2_\circ}^{Q^2_{2}=E_e^2} && \nonumber\\
\frac{d^2\sigma_1}{d\theta_{q1} dQ_{1}^2}  \frac{d^2\sigma_2}{d\theta_{q2} dQ_{2}^2} d\theta_{q1} d\theta_{q2} dQ_{1}^2 dQ_{2}^2 \;\;.&&
\label{eq:xsec_integral}
\end{eqnarray}

Also, the azimuthal angular distance of the two quarks needs to be within the angular interval which
we estimate by reducing the cross-section by
\begin{equation}
\epsilon_\phi = \frac{20\,\rm{deg}}{360\,\rm{deg}} \;\;.
\end{equation}

We solve (\ref{eq:xsec_integral}) numerically for two electrons with energy $E_e = 60$~GeV, 
$Q^2\ge 3.5$~GeV$^2$ and find
\begin{equation}
\sigma^2 = \epsilon_\phi \cdot \tilde{\sigma}^2 = 1.5\cdot 10^{-76} \rm{m}^4 \; .
\label{eq:squared_cross_section}
\end{equation}

\section{Transverse momentum \label{sec:repulsion}}

Depending on their initial color states the two scattered quarks will either repel or attract each other.
Here we are interested in repulsion leading to quarks with substantial transverse momenta with 
respect to the original quark directions after the deep inelastic scattering processes.

As can be seen from the repulsive potential in Fig.~\ref{fig:potential}b, the gain in transverse 
momenta will be largest for two quarks which were initially at the smallest distance from one another.
In order to identify transverse momenta originating from repulsive effects we require them to exeed
transverse momenta naturally arising from Fermi motion.

To estimate the effect, we perform a simulation of the repulsive force between the two quarks
using a semi-classical approach.
The applied model corresponds to a classical interpretation of the repulsive strong potential. It assumes the validity of Newtonian mechanics in the laboratory frame in which two quarks at positions $\vec{r}_1$ and $\vec{r}_2$  experience a Coulomb-like repulsive force as defined by the gradient of $V_{rep}$. It reads
\begin{equation}
\vec{F}_{rep}=-\nabla V_{rep}(r)=\frac{\alpha_s \hbar c}{3\cdot|\vec{r}_2-\vec{r}_1|^3}\cdot\left(\vec{r}_2-\vec{r}_1\right)~.
\end{equation}
The color factor $1/3$ for quark-quark color sextets is incorporated.
Trajectories are computed relativistically implementing Euler's method noting that 
\begin{equation}
\vec{F} = \frac{d\vec{p}}{dt}~~~.
\end{equation}
The repulsive force between the two quarks is calculated for a fixed time. 
Subsequently, its value is used to update the momenta through the addition of the infinitesimal change in momentum $d\vec{p} = \vec{F}\cdot dt$. 
Thus, the quark's positions at the next time step $t \rightarrow t + dt$ are given by
\begin{equation}
\vec{x}_{t} \rightarrow \vec{x}_{t} + \vec{\beta}\, c\, dt
\end{equation}
with $\vec{\beta} = \vec{p}c/E$.
Initial conditions for the computation of the trajectory correspond to the 4-momenta and relative positions 
of the two quarks, while all other possible interactions of the two outgoing quarks are neglected.
To account for possibly large relative changes in the momentum at one iteration step, the step-size is initially set to $dt=10^{-34}~\rm{s}$. It is increased by a factor of 10 every 1000 iterations until $T=\sum_{i=1}^{}dt_i=10~\rm{ns}$, equivalent to $\approx 3$~m of flight distance at the speed of light, is reached.
For two up-quarks at an initial transverse distance $\vec{r}_2-\vec{r}_1 = 1~\rm{fm}\cdot\vec{e}_x$ and parallel momentum along the z-axis $\vec{p}_1=\vec{p}_2=p_z\cdot \vec{e}_z$, the relative transverse momentum with respect to the total momentum grows from 0 to 157 $\rm{MeV}/c$ after 10 ns.

\begin{figure}[htb]
\centering
\includegraphics[width=0.5\textwidth]{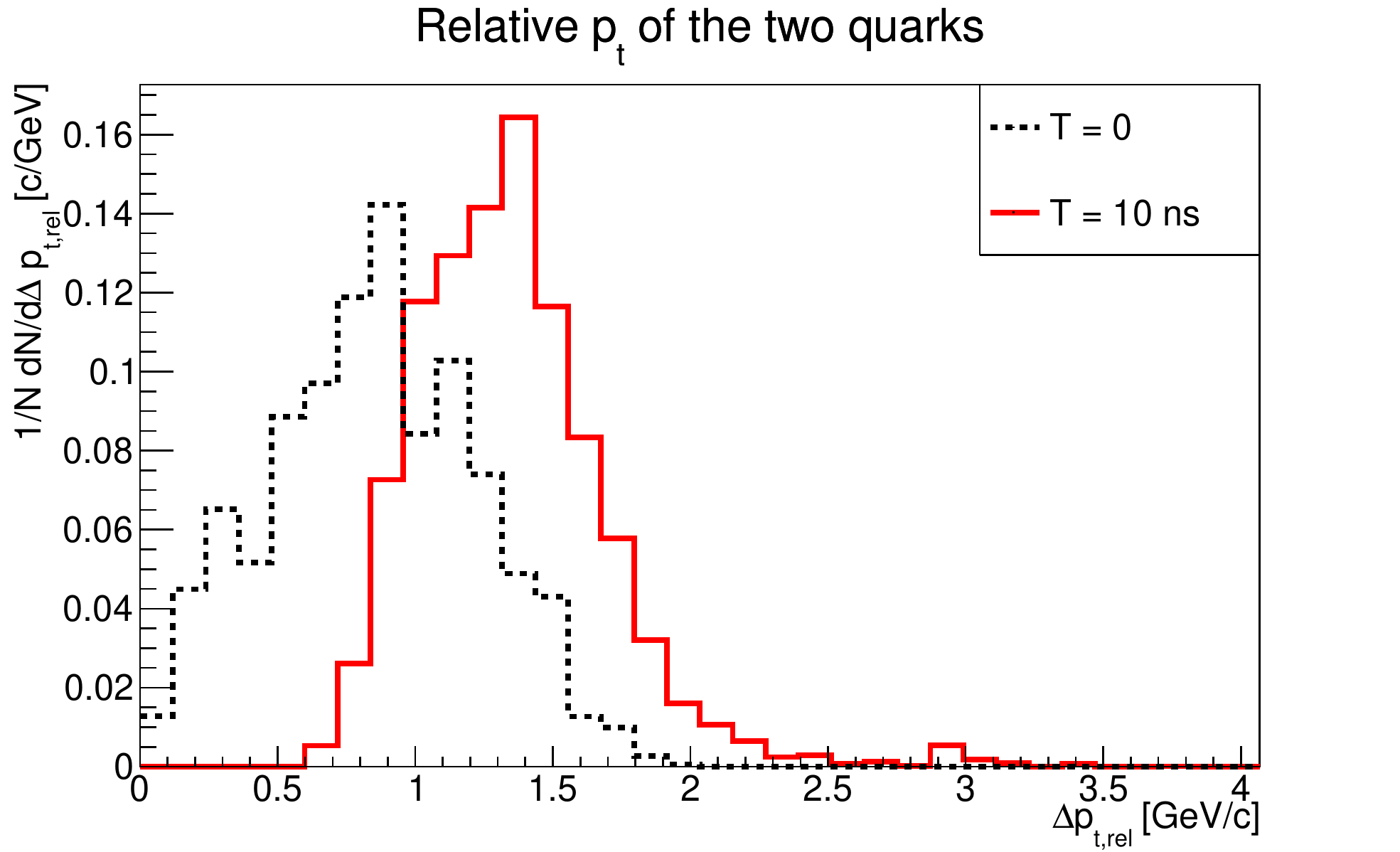}
 \caption{Relative transverse momentum $p_t$ of quarks before and $10 \rm{ns}$ after the repulsive
quark-quark interactions.
}
 \label{fig:repulsion}
\end{figure}

Figure \ref{fig:repulsion} summarizes the study for 10,000 simulated quark-quark interactions. It illustrates that, in general a significant, amount of relative transverse momentum can be gained if the repulsive force repels two otherwise free quarks from 
one another.
For the computation of one entry in these histograms, two of the aforementioned deep inelastic scatterings (DIS) occur within a sphere of 0.5 fm in radius. DIS generating electrons have 60 GeV energy. The angle between the two electron beams amounts to 1 degree. For each DIS individually, both up-type ($m_u=2.3~\rm{MeV/c^2}$) and down-type  ($m_d=4.8~\rm{MeV/c^2}$) quarks are randomly picked in the proton as the collision partners. In agreement with the Fermi motion, their intrinsic momentum components are randomly chosen between $\pm100/\sqrt{3}~\rm{MeV/c}$. Hence, the initial four momenta $p_e$ and $p_q$ of the electrons and quarks are fixed. Subsequently, the kinematics of the outgoing electron and quark are set, such that both the energy-mass and the conservation of 4-momentum is guaranteed
 \begin{equation}
 p_e + p_q ~= ~p_{e'}+p_{q'}~~~.
 \label{eq:4Momentum}
 \end{equation}
Consequently, the eight parameters in $p_{e'}$ and $p_{q'}$ are constrained sixfold. The remaining two degrees of freedom are chosen to be the relative amount of energy $\rho_E$ of the outgoing electron compared to the according summed initial values, and the relative momentum $\rho_z$ along the z-axis:
\begin{equation}
\rho_{E}~=~\frac{E_{e'}}{E_{e}+E_{q}}~~~,
\end{equation}
\begin{equation}
\rho_{p_z}~=~\frac{p_{z, e'}}{p_{z, e}+p_{z, q}}~~~.
\end{equation}
$\rho_{e}$ and $ \rho_{p_z}$ are chosen randomly under consideration of both the energy-mass relation and the conservation of four-momentum\footnote{The choices of $\rho_E$ and $\rho_{p_z}$ are correlated. For instance, $E_e^2-m_e^2 c^4\geq p_{z, e}^2 c^2$ (similarly for the quark) needs to be ensured.}. 

Afterwards, the kinematics of the scattering is fully determined analytically. Initial configurations are rejected from the simulation if the quarks' angular distances exceed $10$~deg. 
The squared momentum transfer $Q^2=-\left(p_e-p_{e'}\right)^2$ and the Bjorken-x $x=Q^2/\left(2\cdot m_p\cdot\left(E_{e}-E_{e'}\right)\right)$ are input to a probabilistic weight $\omega$ given by the product of the differential cross-section (see eq.~\ref{eq:cross section}) representing the likelihood of a generated kinematic configuration:
\begin{equation}
\omega ~=~\frac{F_2\left(x_1, Q_1^2\right)}{x_1 \cdot Q_1^4}\cdot\frac{F_2\left(x_2, Q_2^2\right)}{x_2 \cdot Q_2^4}
\end{equation}
We find that the distributions of the signal events from repulsive interactions 
can be distinguished well from events without such interactions (see Fig.~\ref {fig:repulsion}).

\section{Event rate for two simultaneous interactions with a single iron nucleus}

\begin{figure}[htb]
\centering
 \includegraphics[width=0.3\textwidth]{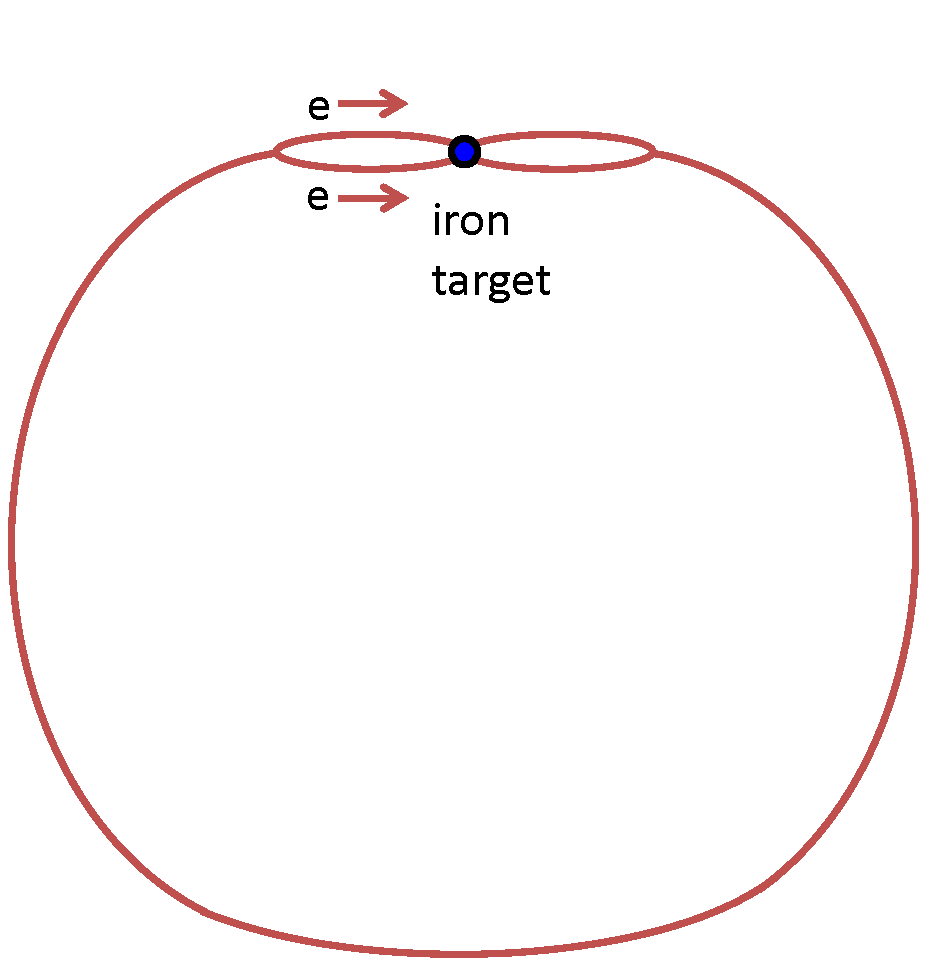}
 \caption{
Storage ring with the electron beam being split before the iron target, and fused after the target.
}
 \label{fig:storage}
\end{figure}

Here, we assume an electron storage ring directing two beams on a single iron target (Fig.~\ref{fig:storage}).
The experimental conditions of the beam and iron target are summarized in Table \ref{tab:parameter}.
A part of the assumed numbers are obviously too large for current technologies.
In the following section will therefore discuss dependencies on different variables and
possibilities to relax the technical requirements.

\begin{table}[htb]
\begin{center}
\caption{\label{tab:parameter} Electron beam and iron target parameters.}
\begin{tabular}{lll}
\\
\hline
\\
Electron beam energy            & $E_e=60$~GeV \\
Radius: electron storage ring   & $R=4.4$~km \\
Deep inelastic scattering       & $Q^2\ge 3.5$~GeV$^2/c^2$ \\
Electrons per bunch             & $N_e=2\;10^{14}$ \\
Electron bunch length           & $l_b=100$~nm \\
Electron bunch area             & $A_b=100$~nm $\times$ $100$~nm \\
Electron bunch distance         & $\delta_b=1\, \mu$m \\
Length of iron target           & $l_\tau=1\, \mu$m \\
\\
\hline
\end{tabular}
\end{center}
\end{table}

The two electron interactions with the same iron nucleus should take place within a 
time frame of about $\Delta t=1$ fm/c.
With two bunches of length $l_b=10^{-7}$~m, the number of electrons appearing simultaneously
at the iron nucleus within $\Delta t$ reduces the effectiveness of one bunch crossing:
\begin{equation}
\epsilon_{\Delta t} = \frac{1 \rm{fm}}{l_b} = 10^{-8}
\label{eq:deltaT}
\end{equation}
The number of target nuclei in the beam of size $A_b$ amounts to (Fig.~\ref{fig:cube}c)
\begin{eqnarray}
N_\tau &=& \frac{N_A\; \rho}{A} \, A_b \, l_\tau  \nonumber \\
       &=& \frac{6\, 10^{23}\; 7.8(\rm{g/cm}^3)}{56\rm{g}} \, (100\times 100)\rm{nm}^2 \times 1\mu\rm{m}  \nonumber \\
       &=& 8.4 \cdot 10^8
\end{eqnarray}
where $N_A$ is the Avogadro constant, A is the molar mass of iron, and $V = A_b \, l_\tau$ is the target volume.
The time for a bunch turn is:
\begin{equation}
T = \frac{2\pi\,R}{c} 
\end{equation}
Here, $R$ denotes the radius of the storage ring.
The number of bunches in the storage ring is
\begin{equation}
N_b=\frac{2\pi\,R}{\delta_b+l_b} \;\;.
\end{equation}
The number of bunches crossing per second amounts to:
\begin{equation}
\dot{N}_b = \frac{N_b}{T} =  \frac{c}{\delta_b+l_b} = 2.7 \cdot 10^{14} \frac{1}{\rm{s}}
\end{equation}
With the electron charge $q_e=1.6 \cdot 10^{-19}$~C, the electron current in the collider would 
then be at the extreme level of
\begin{equation}
I_e = q_e \cdot \dot{N}_b \cdot N_e = 8.7 \cdot 10^{9} {\rm{A}} \;.
\end{equation}
The total number of electrons on target (EOT) per year ($1\,\rm{year}\approx \pi \, 10^7$ seconds) amounts to:
\begin{equation}
EOT = \dot{N}_b \cdot N_e \cdot (1\,\rm{year}) = 1.7 \cdot 10^{36}
\end{equation}

The event rate $\dot{M}_{e}$ for single electron beam scattering off the iron target is calculated
from the luminosity $(\dot{N}_b \; N_e / A_b) \cdot N_\tau$ and the integrated cross section 
(\ref{eq:cross section}):
\begin{eqnarray}
\dot{M}_{e} &=& \frac{\dot{N}_b \cdot N_e}{A_b} \cdot N_\tau \cdot \sigma  
\end{eqnarray}
Correspondingly, the event rate $\dot{M}_{ee}$ for simultaneous deep inelastic scattering
processes with similar directions of the scattered quarks using the product of the cross sections
(\ref{eq:squared_cross_section}) amounts to
\begin{eqnarray}
\dot{M}_{ee} &=& \frac{\dot{N}_b \cdot (N_e/2) \cdot \left(\epsilon_{\Delta t} \cdot N_e/2\right) 
                 }{A_b^2} \cdot N_\tau \cdot \sigma^2 \nonumber\\
             &=& 3.4 \cdot 10^{-5} \frac{1}{\rm{s}} \; .
\label{eq:event-rate}
\end{eqnarray}
Here the splitting of the electron beam in front of the target is included.
The resulting number of events per year thus amounts to
\begin{equation}
M= \dot{M}_{ee}\cdot (1\,\rm{year}) \approx 1,000\; \frac{1}{\rm{a}} \;.
\end{equation}

\begin{table}[htb]
\begin{center}
\caption{\label{tab:dependencies} Dependencies of the event rate on experimental parameters}
\begin{tabular}{lll}
\\
\hline
\\
Electrons in bunch $N_e$               & $\propto N_e^2$ \\
Area of the bunch $A_b$                & $\propto A_b^{-1}$ \\
Bunch length $l_b$                     & $\propto l_b^{-1}$ \\
Target length $l_\tau$                 & $\propto l_\tau$ \\
Rate dependence on                     & $M(E_e=60~\rm{GeV})$ \\
\hfill electron energy \hfill          & $\approx 3 \cdot M(E_e=30~\rm{GeV})$ \\
Simultaneous electron beams            & $\approx 10^8$ \\
\\
\hline
\end{tabular}
\end{center}
\end{table}

\section{Discussion}

The primary challenge in studying repulsive strong interactions is preparing 
two quarks in a sextet color state, or a quark and an antiquark in an octet color state.
In this work we presented a thought experiment preparing the two quarks by deep inelastic
electron-iron scattering.
The product of the cross-sections of two simultaneous interactions is in the order of $10^{-76}$~m$^4$.
To record $1,000$ events resulting from repulsive interactions, 
$10^{36}$ electrons on target are required which is  
extremely high and leads to very large electron currents in a collider.

Different dependencies to optimize the experimental conditions are presented in 
Table~\ref{tab:dependencies}.
A large improvement could be achieved by improving the beam structure
with respect to the simultaneous arrival of the electrons at the target
nucleus.
If a bunch length of $1$~fm can be achieved, 
the efficiency factor for simultaneous appearance of
electrons on target would increase from $\epsilon_{\Delta t} =
10^{-8}$ to $\epsilon_{\Delta t} = 1$ (eq. (\ref{eq:deltaT})). 
Hence, the number of
electrons $N_e$ per bunch could be reduced from $10^{-14}$ to $10^{-10}$
as the event rate depends quadratically on the number of electrons (eq.
(\ref{eq:event-rate})). Correspondingly, the electron current reduces to
the level of $I_e=10^6$~A.

Obviously we have a long way to go to accelerate particles 
by strong interactions in a controlled way.
It seems like a good point in time to inquire about the interest in the community, and to ask 
interested colleagues from theory and experiments to participate in collecting ideas 
and developing a suitable road map.

\section*{Acknowledgments}

This work is supported by the Ministry of Innovation, Science and Research of the State of North Rhine-Westphalia, and the Federal Ministry of Education and Research (BMBF).



\section*{References}

\begin{thebibliography}{}
\bibitem{arneodo} 
M.~Arneodo, et al., The European Muon Collaboration, Proton and Anti-proton Production in 
Deep Inelastic Muon - Nucleon Scattering at 280 GeV, Z. Phys. C 35 (1987) 433 
\bibitem{eichten_theory} 
E.~Eichten, K.~Gottfried, T.~Kinoshita, K.D.~Lane, T.M.~Yan, Charmonium: The Model,
Phys. Rev. D17 (1978) 3090, Erratum: Phys.~Rev. D21 (1980) 313
\bibitem{eichten_experiment} 
E.~Eichten, K.~Gottfried, T.~Kinoshita, K.D.~Lane, T.M.~Yan,
Charmonium: Comparison with Experiment, Phys. Rev. D21 (1980) 203
\bibitem{quigg}
 C.~Quigg, J.~L.~Rosner, Quantum Mechanics with Applications to Quarkonium,
Phys. Rept. 56, 167 (1979).
\bibitem{bali}
G.S.~Bali, QCD forces and heavy quark bound states, Phys. Rept. 343 (2001) 1
\bibitem{hagler}
P.~Hagler, Lattice QCD calculations of hadron structure: Status and perspectives,
Prog. Theor. Phys. Suppl. 187 (2011) 221
\bibitem{ayala} 
C.~Ayala, P.~Gonzalez, V.~Vento, Heavy quark potential from QCD-related 
effective coupling, J. Phys. G43 (2016) no.12, 125002
\bibitem{erdmann1} 
M.~Erdmann, Investigation of quark - anti-quark interaction properties 
using leading particle measurements in $e^+ e^-$ annihilation, Phys. Lett. B 510 (2001) 29
\bibitem{QCD} 
D.~Griffith, Introduction to Elementary Particle Physics, Wiley-VCH, 2012
\bibitem{sansoni}
A. Sansoni for the CDF Collaboration, Quarkonia production at Fermilab, Nuovo Cim. A109 (1996) 827
\bibitem{braaten} 
E.~Braaten, S.~Fleming, T.C.~Yuan, Production of heavy quarkonium in 
high-energy colliders, Ann. Rev. Nucl. Part. Sci. 46 (1996) 197
\bibitem{bodwin}
G.T.~Bodwin, et al., Quarkonium at the Frontiers of High Energy Physics: A Snowmass White Paper,
arXiv:1307.7425 (2013)
\bibitem{pdg} 
C.~Patrignani, et al., The Particle Data Group, The Review of Particle Physics,
Chin. Phys. C, 40 (2016) 100001
\bibitem{emc} 
J.J.~Aubert, et. al. , The European Muon Collaboration, The ratio of the nucleon structure functions 
$F_2^N$ for iron and deuterium, Phys. Lett. B 123 (1983) 275
\bibitem{erdmann2} 
M.~Erdmann, On the structure of the proton, the photon, and color singlet exchange,
Phys. Lett. B 488 (2000) 131
\bibitem{aid} 
S.~Aid, et al., The H1 Collaboration, Energy flow in the hadronic final state of diffractive 
and nondiffractive deep inelastic scattering at HERA, Z. Phys. C 70 (1996) 609
\end{thebibliography}
\end{document}